\documentclass[conference]{IEEEtran}
\IEEEoverridecommandlockouts
\usepackage{tabularx}
\usepackage{float}
\usepackage{placeins}
\usepackage{cite}
\usepackage{balance}

\usepackage{subcaption}
\usepackage{amsmath,amssymb,amsfonts}
\usepackage{algorithmic}
\usepackage{graphicx}
\usepackage{textcomp}
\usepackage{xcolor}

\def\BibTeX{{\rm B\kern-.05em{\sc i\kern-.025em b}\kern-.08em
    T\kern-.1667em\lower.7ex\hbox{E}\kern-.125emX}}
\begin{document}

\title{Performance comparison of 802.11mc and 802.11az Wi-Fi Fine Time Measurement protocols} 

\author{
\IEEEauthorblockN{Govind Rajendran}
    \IEEEauthorblockA{
        \textit{Arista Networks} \\
        govindrajendran98@gmail.com
    }
     \and 
\IEEEauthorblockN{Kushagra Sharma}
    \IEEEauthorblockA{
        \textit{Salesforce} \\
        kushagrasharma30072001@gmail.com
    }
    \and 
    \IEEEauthorblockN{Vijayalakshmi Chetlapalli}
    \IEEEauthorblockA{
        \textit{Arista Networks} \\
        vijaya.chetlapalli@arista.com
    }
    \and
    \IEEEauthorblockN{Jatin Parekh}
    \IEEEauthorblockA{
        \textit{Tarana Wireless} \\
        parekhj@gmail.com
    } 
    
}

\maketitle
\begin{abstract}
The need for meter level location accuracy is driving increased adoption of 802.11 mc/az 
Fine Time Measurement (FTM) based ranging in Wi-Fi networks. In this paper, we present a comparative study of the ranging accuracy of 802.11mc and 802.11az protocols. We examine by real world measurements the
critical parameters that influence the accuracy of FTM {\it{viz.,}}  channel width, interference, radio environment, and offset calibration. The measurements demonstrate that meter-level
ranging accuracy can be consistently attained in line of sight
environment on 80 MHz and 160 MHz channels, while an accuracy of about 5m
is obtained in non-line of sight environment. It is observed that the 802.11az protocol is capable of providing better accuracy than 802.11mc even in a multipath heavy environment. 

\end{abstract}
\begin{IEEEkeywords}
Indoor Locationing, Wi-Fi Fine Time Measurement, Round trip time, 802.11 mc, 802.11az.
\end{IEEEkeywords}
\section{Introduction}
The rapid adoption of Wi-Fi as an ubiquitous solution for indoor network access cannot be overstated, as much of its rise in popularity is due to its cost effectiveness and ease of deployment. Wi-Fi has cemented its place as the defacto indoor network access mechanism. However, the use of Wi-Fi ranging for indoor locationing use cases has been rather slow, mainly due to security and privacy concerns and lack of support by the client devices.
\par
Traditionally,  indoor locationing services relied upon the use of location based on GPS or Bluetooth and other 802.15 based technologies. The GPS signal reception is poor indoors due to severe signal attenuation and scattering. While solutions that use of Bluetooth beacons for navigation exist, they require additional infrastructure that incurs hardware installation overheads. On the other hand, Wi-Fi infrastructure has ubiquitous presence in enterprises, malls, university campuses, stadiums and other large public arenas where indoor locationing plays an important role.
\par
RSSI based locationing systems were often plagued with inaccurate results,  especially in indoor environments.  The signal strength in indoor environments is severely affected by the environment as the signals experience fast fading, needing custom heuristics and filtering algorithms to improve accuracy. Specifically,  in use cases with IoT infrastructure, RSSI based locationing is not a practical and deployable solution, as the IoT devices are constrained in power and processing capabilities.
\par
This resulted in the formation of the IEEE 802.11mc task force and the eventual development of a ToF (Time-of-Flight) based ranging protocol in 2016, which was intended to overcome the shortcomings of RSSI based solutions. Wi-Fi Fine time Measurement (FTM) introduced as part of the 802.11-2016 standard, has attracted much attention in the recent years. FTM measures the round trip time (RTT) of the signal to calculate the distance between the access point (AP) and the client. It is a better approach than the use of RSSI because it is relatively unaffected by signal strength variation. The FTM protocol also enables the AP to share its GPS coordinates along with the civic location address (which can be set at the time of deployment) with the clients. This reduces the overhead of needing a customized signaling mechanism to share these details, further facilitating accurate indoor locationing. The 802.11az protocol released in 2022 improves upon 802.11mc   to provide better security by encrypting the FTM packets and leverages enhanced PHY features of Wi-Fi 6 to effectively mitigate ranging errors induced by multi-path.
\par
While there is ample literature on the working of the 802.11mc protocol, studies on real-life performance of the protocol are very few. Published results on 802.11az performance are also scarce, as the protocol is still new and is not supported by many clients. The primary contribution of this work is a detailed performance comparison of the 802.11mc and 802.11az protocols through real life experiments. We provide insights into the factors affecting FTM performance in general and demonstrate the performance improvements that can be achieved by 802.11az in particular. 
\section{Related works}
\label{sec: Related Works}
A majority of distance estimation methods using signal strength and RSSI metrics exist involve custom processing and finger printing to measure distance. The support of the FTM, ToF based distance measurement technique circumvents these efforts paving the way for more innovative and accurate indoor locationing applications. 
\par
As with all wireless technologies,  a number of external and hardware calibration factors affect the accuracy of FTM.  Oliveira et al. \cite{FTM_Burst_size} highlights the effect of burst size, a important factor that affects the accuracy of distance measurement using commercially available of-the-shelf consumer grade  mobile devices for the 802.11mc protocol. 
In Ibrahim et al. \cite{FTM_performance}, the authors present the use of Wi-Fi Time-of-Flight measurements to improve Wi-Fi locationing accuracy, the study shows that meter-level ranging accuracy is possible, but can only be consistently achieved in low-multipath environments. Zola et al. \cite{zola2021ieee} studies the effects of non-line-of-sight performance of the 802.11mc, with rigorous COTS device experiments, further augmenting their experimental findings of line-of-sight performance of the 802.11mc protocol. These findings show the performance of the early adoption of the FTM ranging technique. Singh et al. \cite{FTM_11mc_preliminary} highlight the performance of the 802.11mc protocol, the effects of multipath and need for accurately time-stamping the packets and the inherent serious risk of the unencrypted fields that can be exploited, a concern addressed in the 802.11az protocol.
\par
Ma et al. \cite{Wi-Fi_RTT_Trilateration} demonstrate a robust trilateration algorithm and error mitigation mechanism adapted towards reducing the locationing error.  This work further motivates us to dwell deeper into  the performance of FTM based measurements while using the 802.11mc/az protocol and the factors that impact this in real Wi-Fi deployment scenarios. The key parameters that affect ranging performance like heterogeneity in the FTM's operational environment including but not limited to different device manufacturers, spectrum (frequency and bandwidth) and external interference are studied in \cite{han2019smartphone}. The authors also  suggest the use of standard error correction techniques for Wi-Fi FTM based ranging systems.
\par
While even during the time of publication of this report the utilization and adoption of 802.11mc is sparse and its performance comparison with the 802.11az protocol is even more rare, we test the performance of these protocols with expectations to observe better performance with the 802.11az protocol. 

\section{FTM protocol}

\begin{figure}[!t]
    \centering
    \includegraphics[width=0.5\textwidth]{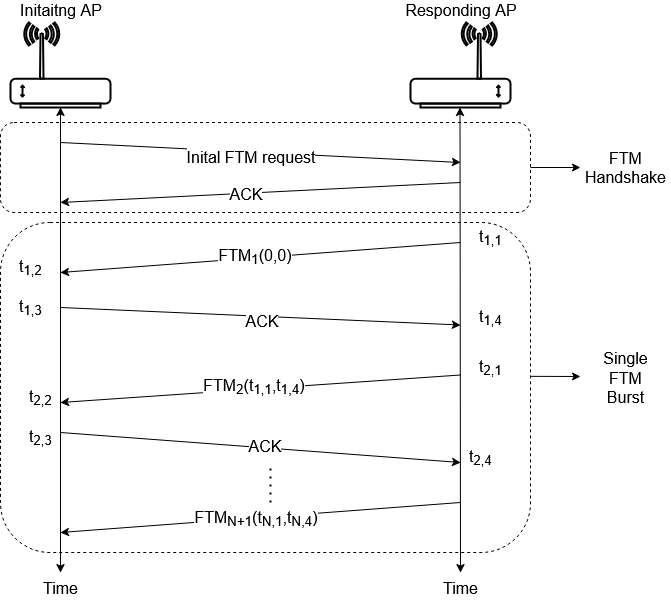}  
    \caption{Illustration of the working of Fine Timing Measurement.}
    \label{fig: FTM_working}
\end{figure}

The 802.11mc and 802.11az protocols are both based on the Time of Flight (ToF) measurement of the radio signal. ToF based distance measurement is based on the simple idea of exchanging the time stamps as illustrated in the Figure~\ref{fig: FTM_working} between the initiator and responder devices. These time stamps are further utilized to find the round trip time (RTT) using (\ref{eq: RTT}). The Distance (D) between the initiating station and the responding AP is calculated as in (\ref{eq: Distance Meaurment}).
\par
In FTM, a session is referred to a specific instance of the FTM protocol between initiator(s) and a responder. It involves a negotiation handshake, exchange of timing measurements, and the eventual termination of the session. The relevant standard \cite{80211azDtandard} can be referenced for a more complete overview of the protocol. During the negotiation phase, the initiating STA requests a preferred periodic time window allocation to execute the FTM procedure. Parameters such as channel width (it is not always required that the FTM session run on the same channel width of the BSS), number of bursts in a FTM session (burst count), FTM frames per burst (burst size) and burst durations are negotiated during the handshake.
\begin{equation}
    RTT= (t_4-t_1)-(t_3-t_2)
\label{eq: RTT}
\end{equation}
\begin{equation}
    D= \frac{(RTT)}{2}\times c
\label{eq: Distance Meaurment}
\end{equation}
It is to be noted that regular CSMA/CA based channel access control rules still apply for FTM frame exchanges. This might cause increased latency in measurements, especially in dense deployment scenarios. Other negotiation parameters, like burst size, and the number of FTM sessions affect the accuracy of the final measurements as we allude to later.
\par 
Since RTT measurements are typically recorded over multiple readings, a variety of approaches can be used to improve the accuracy of the measurements. Simple mean, weighted averages, and median values are typically used. Other complex filtering mechanisms (including removal of outliers)  reduce the mean square error of the distance measurement. It is to be noted that all FTM frame exchanges in a single FTM burst is completed in one transmit opportunity (TxOP).

\subsection{802.11az vs 802.11mc}
\label{11az vs 11mc}
The 802.11az or the Next Generation Positioning (NGP), was ratified in 2022 to improve upon the performance of the 802.11mc FTM ranging mechanism. The 802.11az has many improved performance and management mechanisms, some of which we elucidate in this subsection.
\par
The 802.11az protocol supports ranging with 160MHz channel width, resulting in better accuracy. The accuracy of FTM is directly dependent upon the channel width, the effects of which we allude to in Section~\ref{section: Results}. The 802.11az protocol facilitates AP-AP ranging, which is particularly useful in geolocation of APs to comply with AFC regulations~\cite{AFC}. The AP-AP ranging capability allows for locationing without the additional downtime of switching the initiating AP to STA mode.  A more particularly beneficial feature is the ability to use the MIMO features of Wi-Fi 6 for improved spatial resolution enabling accurate ranging without the need of additional heuristics like~\cite{FTM_Music}. 802.11az improves the scalability of FTM to simultaneously range multiple clients using the trigger based FTM, a feature particularly useful in dense deployment scenarios, as investigated by the work in~\cite{FTM_MUMIMO}. The introduction of long training field (LTF) randomization and 128 bit AES encryption of LTF provide protection from man-in-the-middle attacks. Another important PHY level security feature is the use of 64 bit QAM instead of BPSK. The remainder of the paper presents the results of a detailed comparison of the two generations of the FTM protocol.

\section{Experimental setup}
\begin{figure}[t!]
    \centering
    \includegraphics[width=0.3\textwidth]{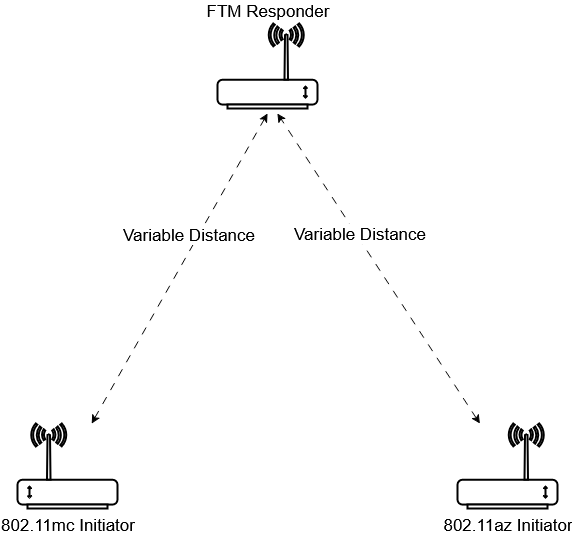}  
    \caption{A illustration of the experimental setup}
    \label{fig: Test_setup}
\end{figure}

In this section, we present a brief description of the experimental setup and scenario. Figure~\ref{fig: Test_setup} is an illustration of the experimental setup and Table~\ref{table: Experimental Componets} highlight the relevant experimental parameters. While most of the existing work in Section~\ref{sec: Related Works} use commercial-off-the-shelf mobile phones to test the performance of FTM, we choose to use an enterprise grade AP both as a responder and initiator. This allows us to tightly control factors like hardware offset calibration and bias, which heavily influence the final outcome of the FTM session. Also, most mobile clients do not support 802.11az yet. As the protocol matures, we expect to see more accurate hardware calibration and 802.11az support in commercial client devices. We run our experiments on the 5 GHz channel 36 in both relatively free and congested scenarios, with varying channel widths.

\begin{table}[t!]
    \begin{tabular}{|c|c|c|}
        \hline
            No. & Component & Type \\
            \hline
            1& FTM Responder& Enterprise-grade (AP), openWrt v 15.5.01\\ \hline
            2.& 802.11mc initiator& Enterprise-grade (AP), openWrt v 15.5.01\\ \hline
            3.& 802.11az initiator& Enterprise-grade (AP), openWrt v 15.5.01\\ \hline
            4.& Operating Channel& Channel 36, with varying channel widths  \\ \hline
            
        \end{tabular}
    \caption{A summary of the relevant experimental components used in our experimentation}
    \label{table: Experimental Componets}
\end{table}

\section{Results}
\label{section: Results}
In this section we demonstrate a detailed performance  comparison of the  802.11az and 802.11mc protocols under different scenarios and evaluate the accuracy of the distance measurements.

\subsection{Effect of Burst size and burst count}

As is the case with all wireless transmission, FTM is also affected by changes in the radio environment. This calls for multiple individual measurements. We vary the Burst Size (number of individual FTM transactions in a burst) and Burst Count (number of bursts in a session), which is currently not possible in COTS devices, as documented by~\cite{FTM_Burst_size}.
\par
\begin{figure}[t!]
    \centering

    \begin{subfigure}{0.8\linewidth}
        \centering
        \includegraphics[width=\linewidth]{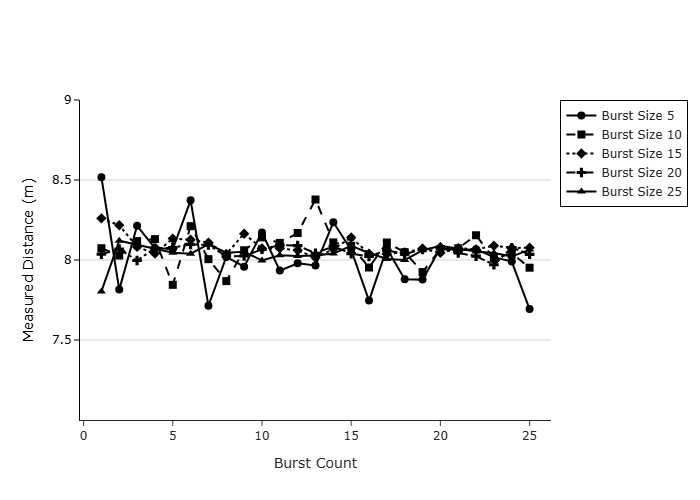}
        \caption{802.11az performance}
        \label{fig:11az_burst}
    \end{subfigure}
    \begin{subfigure}{0.8\linewidth}
        \centering
        \includegraphics[width=\linewidth]{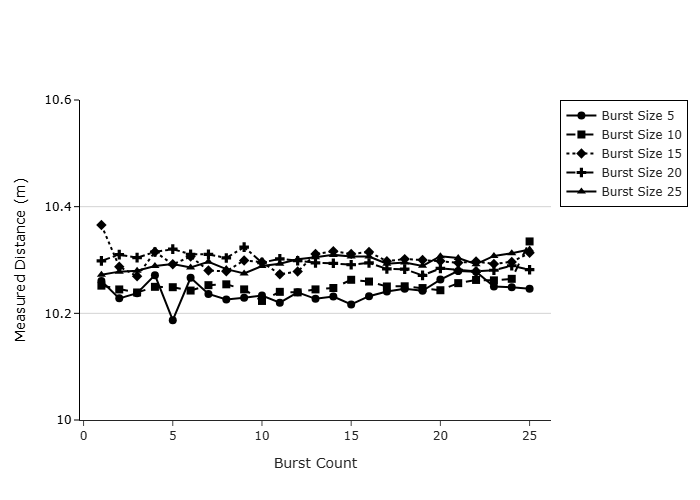}
        \caption{802.11mc performance}
        \label{fig:11mc_burst}
    \end{subfigure}
     \caption{Performance comparison for varying burst sizes and counts on 80MHz and a true distance of 8 meters \textit{(Note that the y-axis is different for the graphs).}}
    \label{fig: burst size and count}
\end{figure}

We compare the performance of 802.11mc and 802.11az in a relatively non-congested RF environment  on channel 36 operating on 80MHz, with minimal external obstructions and reflective surfaces reducing the effect of multi-path. 
The graph in Figure~\ref{fig: burst size and count} shows the performance difference between the protocols with 802.11az demonstrating  accurate, relatively low variance and stable distance measurements as we vary the burst size and burst count. The results are averaged values taken over 30 different measurements with a true distance of 8 meters, and during different times of the day.
\par
It is to be noted that an FTM session with only one frame per burst is not possible as it needs the timestamp from the pervious session to calculate the range. Since, we see that increasing the number bursts and burst count beyond a point is not that useful, we run the all our consecutive experiments with a burst count of 15 and a burst size of 7, totaling out to 105 individual FTM transactions in a session.

\subsection{Effect of Offset calibration}

\begin{figure}[t!]
    \centering
    \includegraphics[width=01\linewidth]{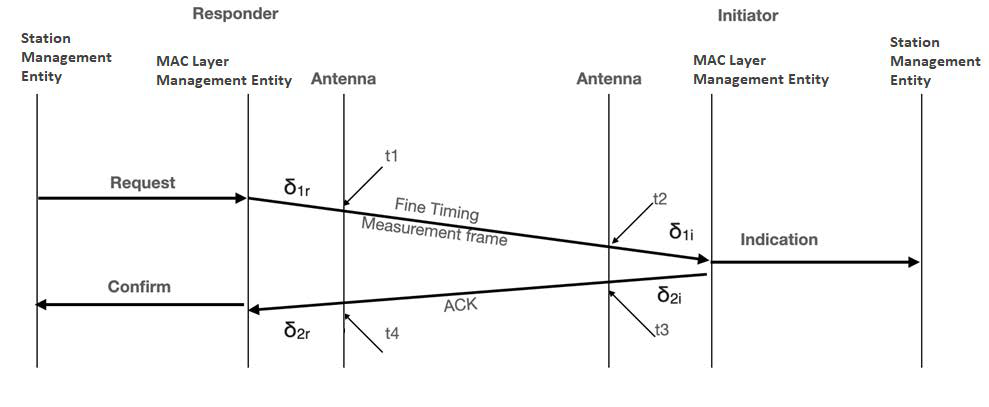}
    \caption{FTM time stamp capture mechanism}
    \label{fig:effect-of-timestamp}
\end{figure}

There is a delay that besets MAC layer time-stamping and the actual transmission of the packets at the physcial layer. These delay values significantly add-up and contribute to the error in FTM measurements. A 1 nano-second error in the RTT value results in a distance error of approximately 0.3 meter. These values often need to be calibrated by the Wi-Fi chipset vendor.  Since we use the calibrated APs as both initiator and responder, the error due to the time stamp delays are negated for on both the initiating and responding APs. The Figure~\ref{fig:effect-of-timestamp} illustrates the need for careful calibration of the values of the delays ($\delta$s).
\par
When the same experiment is repeated by using an uncalibrated responder, large offsets are observed in the measurements. Although there exist some over-the-top solutions for the hardware calibration of the initiator \cite{CalibrationApplication}, it is best done by the chipset vendors. There usually exists a large gap across different chipset vendors, an effect also highlighted by~\cite{FTM_Burst_size}, which affects the performance of the FTM ranging among different OEMs.
We highlight the effect of hardware calibration on the ranging results for one such case in Table~\ref{table: Effect of calibration}.
\par
\begin{table}[t!]
\begin{center}
\begin{tabular}{ |c|c|c|c| } 
\hline
Type& \multicolumn{3}{c|}{Measured Distance (m)}\\
\cline{2-4}
& 20MHz &40MHz &80Mhz\\
\cline{1-4}
Uncalibrated responder& 71.89& 47.89& 66.07\\
\hline
Calibrated responder&10.963&7.453&6.355\\

\hline
\end{tabular}
\caption{Effect of responder calibration on FTM measurements for different channel widths at a true distance of 6 meters.}
    \label{table: Effect of calibration} 
\end{center}
\end{table}

\subsection{Line-Of-Sight Experiments}
\begin{figure*}[!htpb]
\centering
\begin{subfigure}{.48\textwidth}
  \centering
  \includegraphics[width=\textwidth]{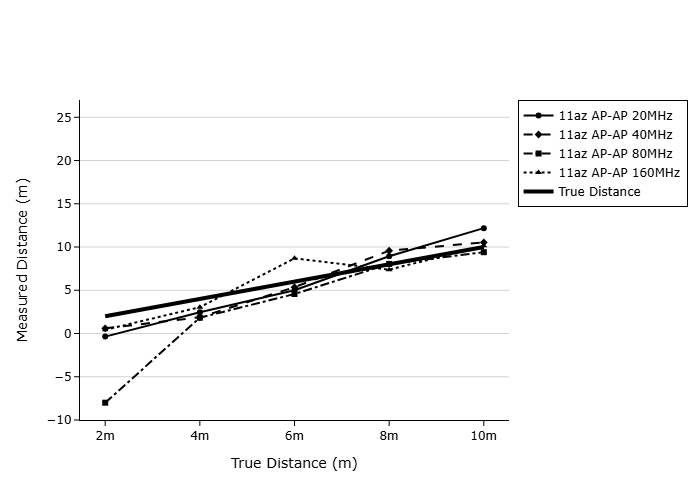}
  \caption{802.11az performance}
  \label{fig:11az}
\end{subfigure}%
\hfill
\begin{subfigure}{.48\textwidth}
  \centering
  \includegraphics[width=1\textwidth]{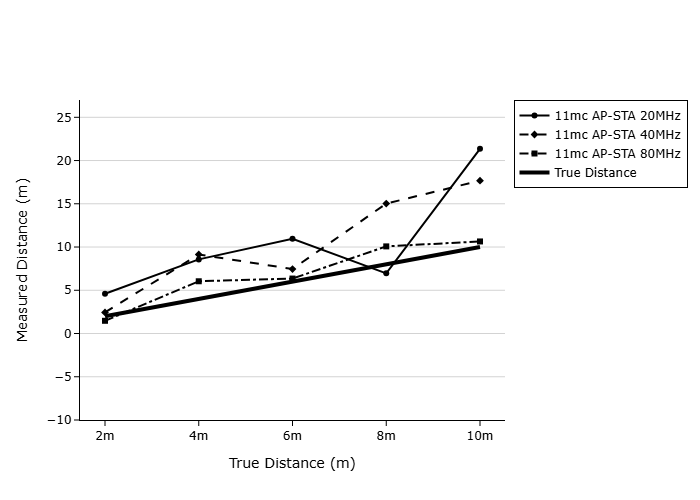}
  \caption{802.11mc performance}
  \label{fig:11mc}
\end{subfigure}
\caption{Comparison of performance of 802.11az and 802.11mc in line of sight conditions with relatively clean RF conditions.}
\label{fig:11azvs11mc}
\end{figure*}

In this subsection, we quantify the performance of the ToF experiments results and highlight the improved performance of the 802.11az over the 802.11mc protocol under line-of-sight conditions (LOS). The graph in Figure~\ref{fig:11azvs11mc} summarizes the performance comparison. We clearly observe more consistent results with the 802.11az under identical test configurations, with improved accuracy results as we go to higher channel widths. The experiments were run in relatively clean RF environments with minimal ambient movements. While the LOS propagation conditions are characterized by none or low multi-path effect, this allows us to clearly witness the effect the channel width has on the performance of FTM measurements. It is also quite evident that the true distance of 2 meter performance is heavily hindered even in the 802.11az experiments. The exact reasons for this performance anomaly is yet to be analyzed.
\par
\begin{figure*}[!htbp]
\centering
\begin{subfigure}{.48\textwidth}
  \centering
  \includegraphics[width=\textwidth]{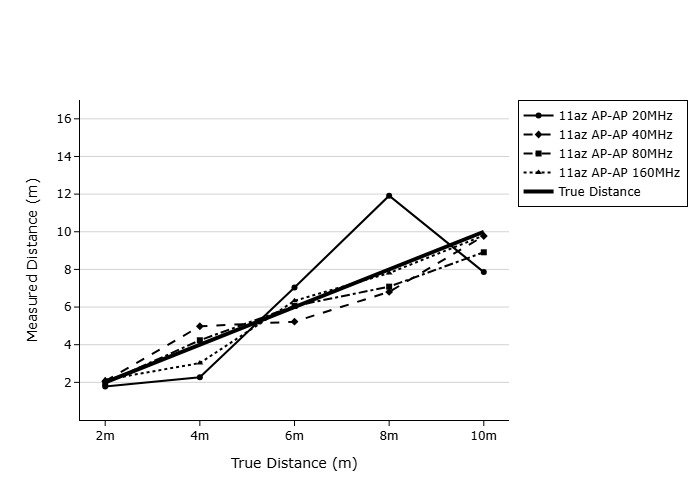}
  \caption{802.11az performance}
  \label{fig:11az_nlos}
\end{subfigure}%
\hfill
\begin{subfigure}{.48\textwidth}
  \centering
  \includegraphics[width=\textwidth]{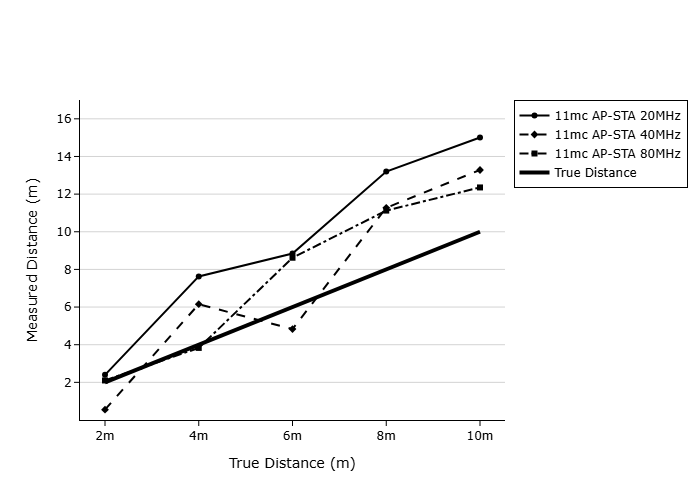}
  \caption{802.11mc performance}
  \label{fig:11mc_nlos}
\end{subfigure}
\caption{Comparison of performance of 802.11az and 802.11mc in line of sight conditions in the presence of congestion and multi-path.}
\label{fig:11azvs11mc_los_dirty}
\end{figure*}

We also compare the performance of both protocols in a congested RF environment, abundant with reflecting surfaces. The graph in Figure~\ref{fig:11azvs11mc_los_dirty} summarizes the better performance of 802.11az over 802.11mc, with majority of the distance measurements being closer to the true distance metric.

\subsection{Non-Line-Of-Sight Experiments}
\begin{figure*}[!ht]
\centering
\begin{subfigure}{.48\textwidth}
  \centering
  \includegraphics[width=\textwidth]{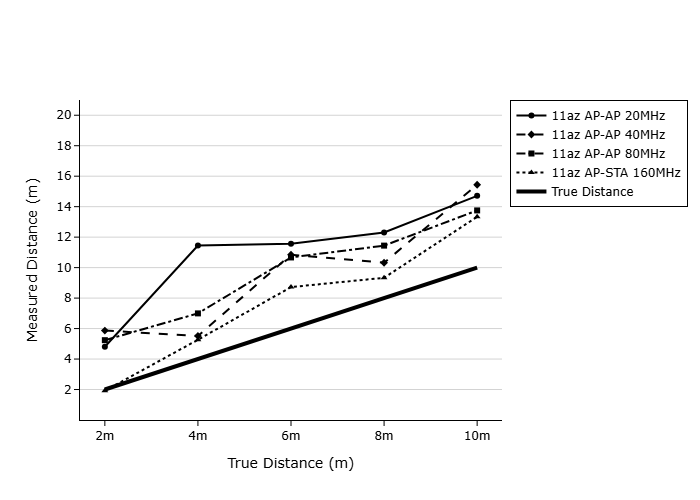}
  \caption{802.11az performance}
  \label{fig:11az_nlos}
\end{subfigure}%
\hfill
\begin{subfigure}{.48\textwidth}
  \centering
  \includegraphics[width=\textwidth]{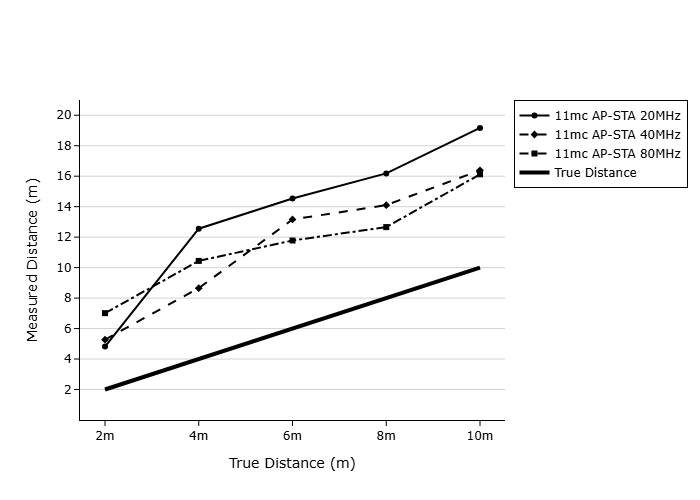}
  \caption{802.11mc performance}
  \label{fig:11mc_nlos}
\end{subfigure}
\caption{Comparison of performance of 802.11az and 802.11mc in non-line of sight conditions with relatively clean RF conditions.}
\label{fig:11azvs11mc_nlosclean}
\end{figure*}

While LOS experiments provide a good baseline for the comparison of the protocols, it is often the case that we need distance estimation in non-line-of-sight (NLOS) conditions, a common occurrence in dense urban indoor deployments. In this subsection, we highlight the improved performance of the 802.11az ranging protocol and how the performance is further improved in NLOS conditions,

The graph in Figure~\ref{fig:11azvs11mc_nlosclean} illustrates the increased performance in a NLOS, when run in relatively clean RF conditions. While we do not observe significant improvements in absolute accuracy, 802.11az still outperforms 802.11mc with respect to the accuracy of the measurements.
\par
\begin{figure*}[t!]
\centering
\begin{subfigure}{.48\textwidth}
  \centering
  \includegraphics[width=\textwidth]{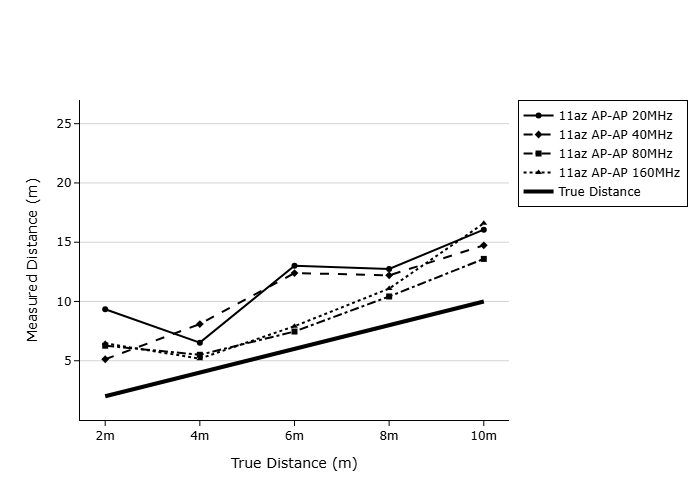}
  \caption{802.11az performance}
  \label{fig:11az_nlos}
\end{subfigure}%
\hfill
\begin{subfigure}{.48\textwidth}
  \centering
  \includegraphics[width=\textwidth]{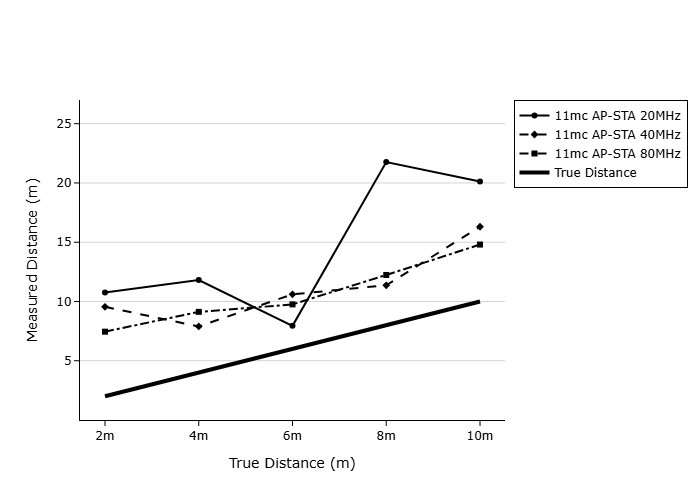}
  \caption{802.11mc performance}
  \label{fig:11mc_nlos_dirty}
\end{subfigure}
\caption{Comparison of performance of 802.11az and 802.11mc in non-line of sight conditions in the presence of heavy congestion and multi-path.}
\label{fig:11azvs11mc_nlos}
\end{figure*}

In our final performance comparison, we test the protocols in congested RF scenarios with many reflecting surfaces and a congested RF environment similar to the LOS experiments. The graph in Figure~\ref{fig:11azvs11mc_nlos} illustrates the comparison. We observe that at higher channel widths (40 MHz and above), 802.11az clearly provides better accuracy than 802.11mc due to the improvements in using MIMO to combat multipath efficiently. 
\par

FTM packet failure rates of were observed to be at 50\%, 37\%, 55\%, 50\% and 65\% for 2m, 4m, 6m, 8m and 10m cases respectively, in the congested RF scenario for 802.11az. In a densely congested scenario, the need for larger channel widths for ranging, might be a challenging deployment consideration that can affect the performance of the FTM protocols.

\section{Conclusion and Discussion}
\label{sec: Conclusion and future Scope}
As the need for accurate locationing and ranging gains importance along side the adoption of Wi-Fi as the de-facto network access mechanism indoors, a scalable, accurate, standardized ranging protocol viz. FTM was introduced in Wi-Fi.
\par
In this work, we evaluate the accuracy and effectiveness of the FTM protocol experimentally, with state-of-the-art commercial enterprise grade Access Points. We observe improved accuracy of the latest FTM protocol version 802.11az. Significant improvements in performance were evident especially in congested and noisy RF environments. However, a sub 1 meter accuracy seems difficult to achieve in challenging RF enviroments. 
\par
Much of the improvement can be realized as we move to higher channel widths, an option that is difficult in dense deployment scenarios. As the technology matures and the need for indoor locationing increases, we hope to see better implementation of the protocol and its wide-spread adoption. As locationing use cases in dense deployments evolve, the use of scarce airtime resources for FTM ranging needs to be optimized intelligently so as not to affect the access performance of the APs.  Careful deliberation and optimized approaches must be used for advantageous utilization of FTM. This paper offers real world measurement insights into the effects of various parameters on FTM ranging accuracy of the latest 11az protocol and opens up avenues for possible improvements. 
\balance
\bibliographystyle{IEEEtran}  
\bibliography{references} 

\end{document}